\documentclass[a4paper,10pt]{article}
\usepackage[utf8]{inputenc}
\usepackage{graphicx}
\usepackage[numbers]{natbib}

\title{21 years of Astronomy at Warwick:\\ celebrating the legacy of Prof.\,Tom Marsh}
\author{}
\date{}

\begin{document}

\maketitle
\vspace{-1.5cm}

\begin{figure}[h!]
\centering
\includegraphics[width=0.6\textwidth]{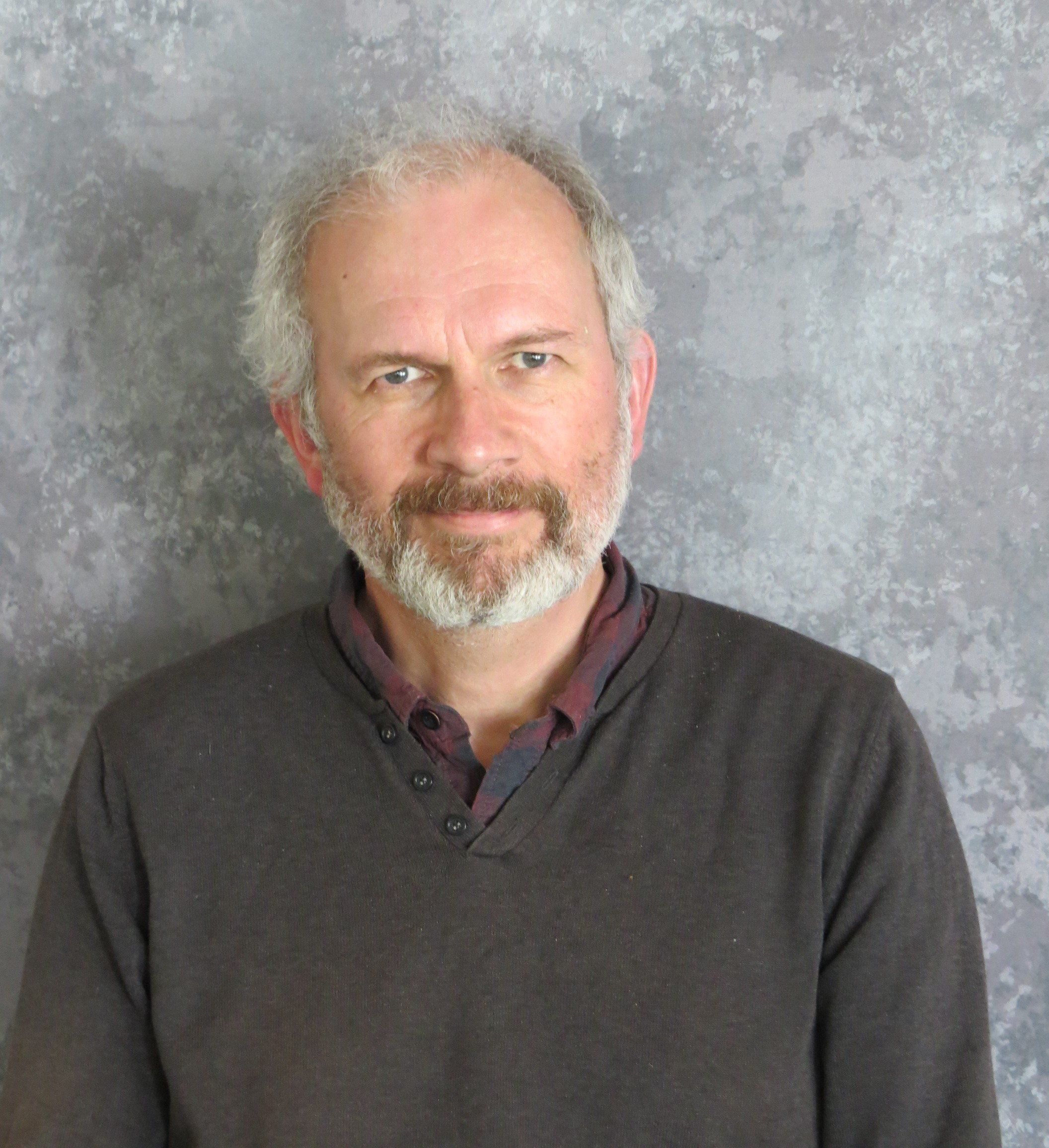}
\label{fig:tom}
\end{figure}

Between the 4th and 6th of September 2024, the Astronomy \& Astrophysics group at the University of Warwick held a meeting to celebrate 21 years of astronomy at Warwick and the scientific legacy of the late Prof. Tom Marsh, the group founder. More than a hundred people attended the meeting, with about half of the attendees being external delegates and coming from as far afield as the USA and South Africa. Tom Marsh moved to the University of Warwick from Southampton in 2003, after the Department of Physics decided to expand the scope of its research. From its humble beginnings with only two staff members, Tom himself and Boris G\"{a}nsicke, one postdoc and a couple of PhD students, the group has now grown to more than 95 members, including 25 staff.

The first two days of the meeting were dedicated to Tom’s scientific legacy, covered by the articles in Section~\ref{sec1}. We heard about Tom’s key role in developing the technique of Doppler tomography, which enabled unique insight into the structure of accreting stellar systems. Tom’s contributions to the field of double-degenerate binaries, one of the major strengths of the Astronomy \& Astrophysics group, were highlighted by many of the speakers. We also heard about Tom’s work in instrumentation, in particular in the development of the state-of-the-art high-speed photometers ULTRACAM, ULTRASPEC and HiPERCAM, which have enabled Warwick to be at the forefront of some fantastic discoveries, such as the white dwarf pulsar AR~Scorpii \cite{Marsh2016}. The talks provided a pleasant mix of scientific overview and testimonial, with many people sharing pictures and anecdotes of happy moments lived with Tom.

The last day of the meeting was focused on the history of the Astronomy \& Astrophysics group at Warwick, as detailed in Section~\ref{sec2}. We were joined by the Vice-Chancellor of the University of Southampton, Professor Mark E. Smith CBE, who was Pro Vice-Chancellor for Research at the University of Warwick when Tom was hired. He explained the background behind the strategic decision of expanding the Department of Physics and its research scope, which ultimately led to Tom’s hiring. Prof.\@ Boris G\"{a}nsicke provided an overview of the group’s past achievements, and Prof.\@ Don Pollaco summarised our present and discussed possible directions the group might take in the future, in particular with the soon to be built Digital Telescope Prototype, awarded £3M on an ERC Advanced Grant. 

The meeting was closed with a “Share a memory” session, where all attendees were invited to share an anecdote or picture related to Tom or Warwick. We heard stories from many, from senior researchers who met Tom as a graduate student to current students who met Tom as a lecturer. Common topics in all cases were Tom’s kindness, humour, and brilliant intellect. Tom was a positive influence to all who met him and will be forever missed by our community, but we can take solace from witnessing his spectacular legacy living on.

\section{The Scientific Legacy of Tom Marsh}
\label{sec1}

\subsection*{Doppler Tomography}
{\large Keith Horne \& Danny Steeghs}
\vspace{3mm} 

One of Tom Marsh's greatest legacies, recognised by his being awarded
The Royal Astronomical Society’s 2018 Herschel Medal, 
is the
pioneering development and extensive application of Doppler Tomography, a method that converts phase-resolved emission-line velocity profiles into 2-dimensional images of the rotating velocity field of a binary star system.
This method enables indirect imaging on micro-arcsecond scales of the accretion flows in cataclysmic variable stars (CVs) and related objects.

\begin{figure}[h!]
\includegraphics[width=\textwidth]{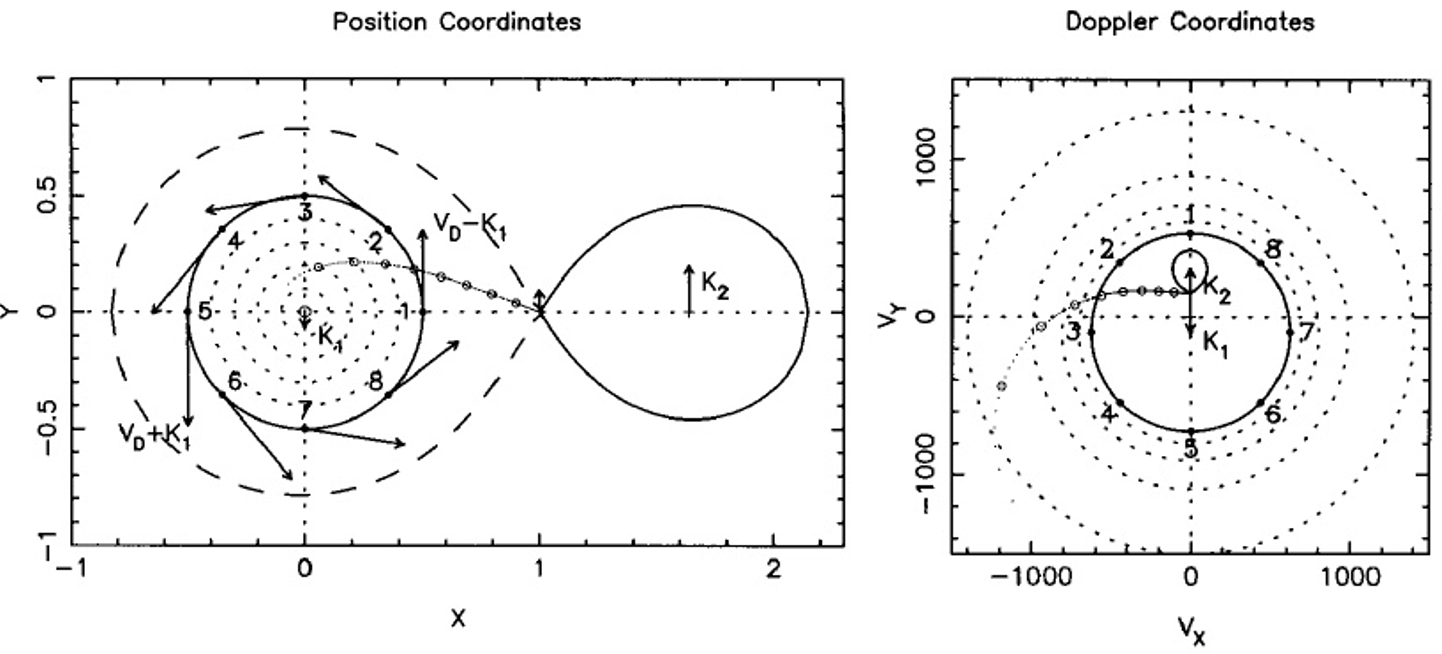}
\caption{Doppler maps of emission-lines in a cataclysmic variable star label each point in the system by its velocity vector ($V_X,V_Y)$ rather than position vector $(X,Y)$ in the rotating frame of the binary.
\label{fig:xy2vxvy}}
\end{figure}

The Doppler Tomography era began in 1983, when Tom was a graduate student in Jim Pringle's group at the Institute of Astronomy in Cambridge.
In the Cambridge Observatory library, some members of the group came across a conference proceedings on medical imaging.
While reading this carefully to understand how medical CAT scans perform the magic of imaging human interiors without surgery, we realised that the exact same method could be applied to the emission-line velocity profiles. 
Each velocity profile is a projection of the velocity field along the line of sight, with different projections at different phases of the rotating binary system.
The result is a 2-dimensional image in Doppler velocity coordinates ($V_X,V_Y)$, whose projections give the velocity profile at different phases (see Fig.\,\ref{fig:xy2vxvy}).

Our first Doppler maps, constructed in 1983 from simulated data, used a Fourier-filtered back-projection (FFBP) algorithm identical to that used at the time for medical CAT scans. 
Doppler maps are just CAT scans of the observed velocity-profile projections.
These early FFBP maps suffered from radial spoke artefacts reflecting the limited number of binary rotation phases available (in turn due to slow CCD readout times).
Tom harnessed the MEMSYS inferrence engine \cite{SkillingBryan1984} to implement 
a maximum entropy regularisation method
similar to that used for eclipse mapping \cite{Horne1985}.
By delivering the smoothest positive image that fits the data, this approach nicely suppressed the radial spokes, and permitted extensions to include orbital modulations \cite{Steeghs2003}, and 3-D mapping \cite{Marsh2022}.
Tom's algorithm \cite{MarshHorne1988} became the workhorse tool for Doppler mapping of the accretion flows in binary systems, which Tom and his collaborators and many others used with great impact.

\begin{figure}[h!]
    \centering
    \includegraphics[width=\linewidth]{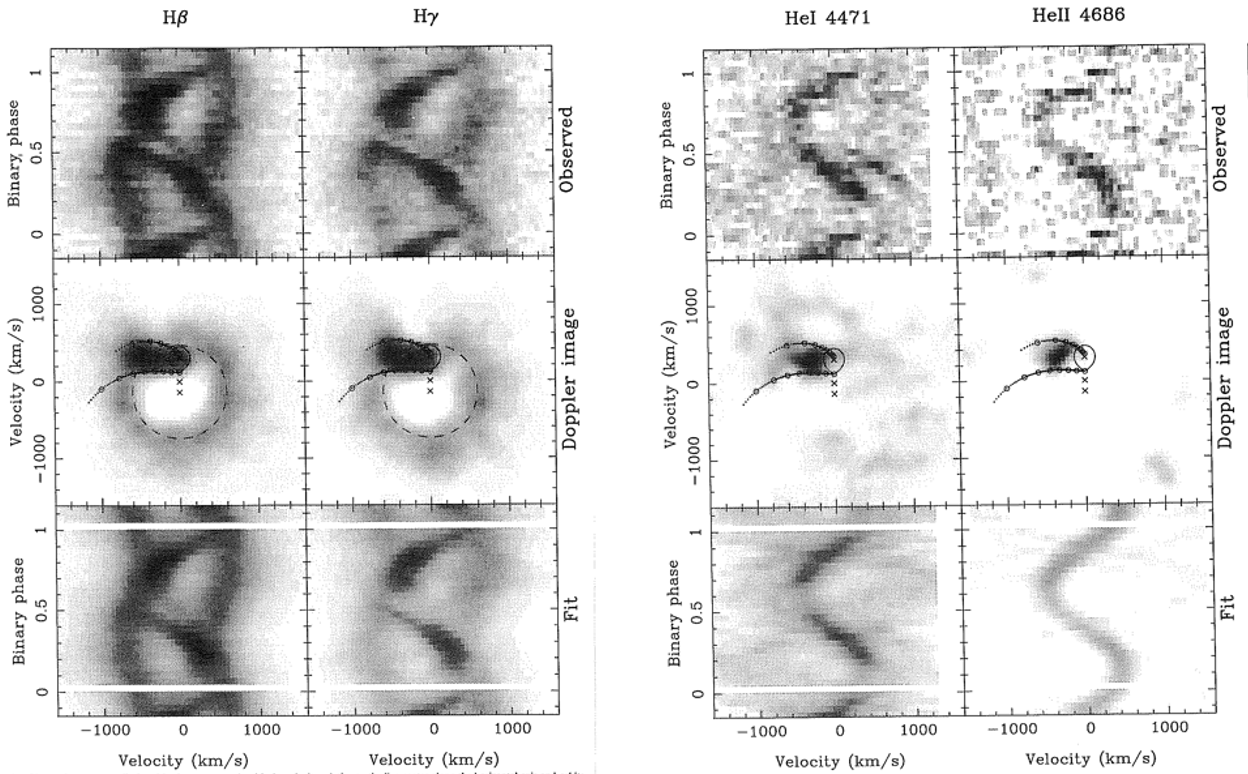}
    \caption{Doppler maps of dwarf nova U\,Gem in four different emission lines, resolving the kinematics and ionisation structure of the accretion flow  \cite{Marsh1990}.
    \label{fig:ugem}
    }
\end{figure}

Quiescent dwarf novae are CVs that display strong double-peaked emission lines arising from gas 
orbiting in the accretion disc around the white dwarf primary.
With contours of constant Doppler shift forming a dipole pattern on the Keplerian disc, 
the observed emission in each velocity bin corresponds to emission in a particular dipole sector \cite{HorneMarsh1986}. 
The double-peaked emission lines project to form rings on the Doppler map (see Fig.\,\ref{fig:ugem}), revealing the radial and azimuthal structure of the line emission.
The typical radial structure scales as $R^{-1.5}$, thus proportional to the Kepler orbit frequency, similarly to the chromospheric emission lines from magnetic activity on rotating stars \cite{HorneSaar1991}.
The ``S-wave'' emission-line component that snakes back and forth between the twin peaks arises from the irradiated Roche-lobe-filling donor star and/or the ballistic gas stream emerging from the nozzle flow through the inner Lagrangian point.

\begin{figure}[h!]
    \centering
    \includegraphics[width=\linewidth]{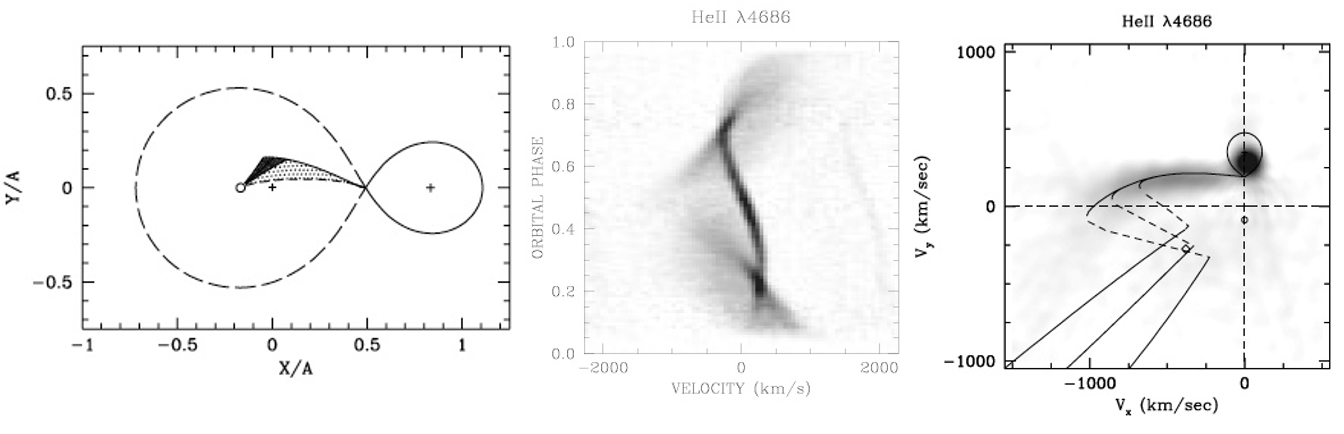}
    \caption{Doppler map of HU\,Aqr \cite{Schwope1997} 
    highlighting He\,II emission on the irradiated face of the donor star Roche lobe, the ballistic stream emerging from the inner Lagrangian point, being stripped down to form a magnetic curtain that funnels the accretion flow onto a small X-ray shock on the white dwarf surface.
    \label{fig:huaqr}}
\end{figure}

Doppler tomography excels at stacking up weak signals to detect the projected orbit velocity $K_2$ of the secondary stars in accreting binaries, 
enabling the determination of masses for the accreting white dwarfs, neutron stars, and black holes, e.g. \cite{Marsh1994, Harlaftis1996, Casares1997}. 
Doppler tomograms also revealed new phenomena, such as slingshot prominences \cite{Steeghs1996} and spiral waves in the disc \cite{Steeghs1997}.
In the AM\,Her systems (a.k.a. Polars), the gas stream is stripped down and captured by the synchronously locked white dwarf's magnetosphere,
feeding a magnetic funnel onto the white dwarf.
X-rays from a standoff shock at the base of the funnel irradiate the incoming accretion flow and the inner face of the donor star. 
These structures are clearly delineated in Doppler maps (see Fig.\,\ref{fig:huaqr}).
In the Intermediate Polars, the inner disc is captured by the white dwarf magnetosphere, giving line profiles that vary on both the binary orbit period and the shorter white dwarf spin period. 
Tom developed stroboscopic Doppler mapping \cite{MarshDuck1996} to construct spin-resolved movies as the rotating irradiation pattern sweeps across the structures of the accretion flow.
Doppler maps of the magnetic propeller
system AE\,Aqr \cite{Wynn1997}
track the ejection of the gas stream by the rapidly spinning white dwarf magnetosphere.

Offshoots of Doppler Tomography have migrated into other areas of astrophysics.
Zeeman-Doppler imaging of stellar surfaces \cite{Donati1997} interprets phase-resolved absorption-line profiles with bumps that shift from blue to red across the line profile, disappearing and reappearing at the stellar limb as the star rotates.
In CVs, Roche tomography \cite{RuttenDhillon1994} similarly maps starspots on the donor star in AE\,Aqr, resolving the ``nose'' of the Roche lobe \cite{Hill2016}, showing that the star is not tidally locked, and even detecting differential rotation with latitude \cite{Hill2014}.
Similar methods recording the silhouettes of transiting exoplanets \cite{Cameron2010}
can measure the obliquity and nodal precession of exoplanet orbits \cite{Johnson2015,Watanabe2022},
and enhance the detection of molecules in exoplanet atmospheres
\cite{Matthews2024}.
Reverberations in the emission-line profiles of active galactic nuclei are used to 
measure their black hole masses and construct
2-dimensional velocity-delay maps 
of their broad-line regions 
\cite{Horne2004,Horne2021}.
Thus the impact of Tom's pioneering development of Doppler Tomography extends well beyond its original focus on accreting binaries.


\subsection*{Double-degenerate binary systems}
{\large Gijs Nelemans \& Kevin Burdge}
\vspace{3mm} 

Double white dwarfs are one of the most common types of outcome of binary evolution and, at the same time, are important for a number of fields within astrophysics. They are key for understanding binary evolution at low mass, they are the most abundant population of mHz gravitational wave (GW) sources and when they get so close that mass is transferred from one WD to another (at staggering orbital periods of mere minutes), they can give rise to spectacular objects and transients, including type Ia supernovae.

However, it has been hard to find them. In the early 1990s only one close double white dwarf was known. Tom made a lasting contribution to the field by his strategic search for binaries using radial velocities to discover the movement of (one of) the white dwarfs. His brilliant idea was to target low-mass white dwarf, which theoretically can almost only be formed in binaries, and it was successful \cite{1995MNRAS.275L...1M,1995MNRAS.275..828M}. Later Tom became heavily involved in the ESO supernovae type Ia progenitor (SPY) survey that discovered many more double white dwarfs \cite{2020A&A...638A.131N} and most recently started executing his plan from much earlier to use Gaia measurements to select white dwarfs with anomalous brightness as an indicator of binarity. Again, a technique that has been highly successful \cite{2024MNRAS.532.2534M}. All in all, many tens of double white dwarfs have now been discovered through radial velocity measurements, thanks to Tom. An ADS search for ``double white dwarf'' in full text shows a 10-fold increase in the number of articles since 1995 (while articles with ``binary star'' increased only by a factor $<2$).

Building on his groundbreaking early-2000s work on the periodic X-ray variables V407 Vul \cite{Marsh2002} and HM Cancri \cite{2004MNRAS.350..113M}, Tom pioneered the idea of using photometric variability to identify double white dwarfs. He developed the theoretical framework for the direct impact model that explains the modulation in V407 Vul and HM Cancri, and shortly thereafter observed V407 Vul with ULTRACAM, revealing detectable optical periodicity \cite{Marsh2005}.

Following the discovery of the first eclipsing double white dwarf, NLTT~11748, Stephen Parsons and Tom led the identification of CSS~41177 — an eclipsing double helium-core white dwarf binary discovered in the Catalina Sky Survey \cite{Parsons2011}. Their work laid the groundwork for the highly successful searches conducted by modern synoptic time-domain surveys such as the Palomar Transient Factory (PTF), the Zwicky Transient Facility (ZTF), the Asteroid Terrestrial-impact Last Alert System   
 (ATLAS), and the Transiting Exoplanet Survey Satellite (TESS). This era of discovery will culminate with forthcoming powerful facilities, including the Roman Space Telescope and the Vera C. Rubin Observatory.

We will further highlight the impact of Tom’s work on double white dwarfs in four themes:

\subsubsection*{Binary evolution}
Already the first set of double white dwarfs discovered by Tom and his co-workers showed that in most systems the two white dwarfs had similar masses \cite{2002MNRAS.332..745M}. This was in contrast with the models in which the younger white dwarf was significantly lower in mass for most of the population that experienced two common-envelope (CE) phases, or a bit higher in mass for a smaller part that experienced only one CE, see \cite{2002MNRAS.332..745M}. It led to the proposal to describe the first CE in terms of an angular momentum balance instead of an energy balance, which gives acceptable results \cite{2000A&A...360.1011N}. Recently, it has been found that CE may be avoided completely in the first phase of mass transfer, e.g. \cite{2023A&A...669A..45T}, but that produces double white dwarfs with specific mass ratios, in contrast to the observations. This conundrum may be related to the discovery that many binaries that have experienced one phase of mass transfer have eccentric orbits, which is not expected, e.g. \cite{2018A&A...620A..85O}.

\subsubsection*{Gravitational waves}
Despite the puzzles surrounding the evolution of binaries, the work by Tom and others has firmly established that there is a significant population of close double white dwarfs \cite{1999MNRAS.307..122M}. They will be the most numerous sources for a mHz GW detectors, such as the LISA mission, e.g. \cite{2001A&A...375..890N,2011CQGra..28i4019M}. In fact, the expected number of sources in the LISA band is huge ($\sim 10^8$), with several (tens of) thousands expected to be individually measured, see \cite{2023LRR....26....2A}. This dataset will be fantastic to constrain the double white dwarf merger rates as function of mass, learn about binary evolution, learn about the physics of the onset of mass transfer \cite{2004MNRAS.350..113M} and even tell us about the structure of the Milky Way \cite{2019MNRAS.483.5518K}.

\subsubsection*{Type Ia supernovae} Tom laid the groundwork for campaigns to identify double white dwarfs, and recent discoveries increasingly link these systems to Type Ia supernovae. In particular, studies of hypervelocity stars \cite{Shen2018, ElBadry2023} strongly suggest that close double white dwarf binaries are their origin, with one white dwarf exploding as a Type Ia supernova. One of Tom’s most recent ideas – a survey for double-lined white dwarfs—led to the discovery of a super-Chandrasekhar white dwarf pair only 49 parsecs away \cite{2024MNRAS.532.2534M}, demonstrating that Type Ia progenitors may be more common than previously thought. Most recently, \cite{Chickles2024} and \cite{Kosakowski2024} reported a LISA-detectable Type Ia progenitor at 500 pc, using the photometric techniques Tom pioneered. Future photometric surveys, such as Rubin and Roman, along with LISA, promise to refine our understanding of these critical standard candles, and Tom had the insight to guide the field in this direction at an early stage.

\subsubsection*{AM CVn binaries}
Tom pioneered the study of AM CVn–type binaries, in which a white dwarf accretes helium-rich matter from a degenerate companion. His early work on Doppler tomography proved instrumental in revealing the accretion structure of these ultracompact systems. He led the initial efforts to characterize the kinematics of accretion in the AM CVn system GP Com \cite{Marsh1999} and played a major role in subsequent studies of these objects.
Thanks to modern time-domain surveys, the detection of AM CVns—via both outbursts and periodic photometric variability—has entered a golden era. Tom contributed significantly to characterizing the 10.3-minute AM CVn system ES Ceti, which displayed partial eclipses of its accretion disk. Recently, several more systems at similarly ultrashort periods have been reported, an exciting development given that most of these sources lie within the LISA gravitational wave band. As upcoming surveys reveal a growing sample of AM CVns, there will be immense opportunities to build on Tom’s legacy of understanding their unique accretion behavior, offering insights that extend across the broader population of interacting binary stars.


\subsection*{AR Scorpii}
{\large David Buckley \& Ingrid Pelisoli}
\vspace{3mm} 

Despite its discovery as a moderately bright variable star over a hundred years ago, using photographic plates, the true nature of AR Scorpii was only revealed in 2016, in a Nature paper led by Tom \cite{Marsh2016}. Photoelectic photometry in the 1970s had resulted in its classification as a $\delta$-Scuti pulsating variable star with a period of a few hours \cite{Satyvaldiev1971}, but observations by amateur astronomers revealed that, apart from a 3.6 h quasi-sinusoidal period, there was shorter timescale high amplitude variability, inconsistent with a $\delta$-Scuti. This information was conveyed to Boris G\"{a}nsicke and Tom, who then observed it with ULTRACAM, revealing a strong $\approx 2$~min periodicity. The rest, as they say, is history.

The unique nature of AR Sco was revealed from a series of multi-wavelength observations, from the UV (HST+COS), optical (ULTRACAM), infrared (HAWK-I) and radio (ATCA), plus other archival data. Additional spectroscopy showed that AR Sco is a white dwarf - red dwarf detached binary, with a 3.6 h orbital period. But the key discovery was the detection of two coherent periods, at 117.1 and 118.2~s, the former identified as the spin period of the white dwarf and the latter being the synodic, or beat, period with the orbit. The latter was seen to dominate the power spectrum at longer wavelengths. These modulations are at a very high amplitude, reaching a pulse fraction of 90\% in the UV, and are also seen in the harmonics. 

Combining all of the available data in a spectral energy distribution (SED) plot showed that all the multi-wavelength emission, from radio to X-rays, could be characterized by a combination of non-thermal (presumably synchrotron) emission combined with the emission from the red dwarf companion star. Due to the low temperature of the white dwarf, it was not directly detected at the time. The non-thermal emission was seen to be modulated on both the spin and beat periods, consistent with magnetic dipole emission from a strongly magnetized fast rotating white dwarf orbiting with its companion. The confirmation of this interpretation came from the determination of the period derivative from 7 years of archival Catalina Real Time Survey (CRTS), implying that the white dwarf spin is slowing down and will synchronize on a timescale of $10^7$ years. Dipole emission based on a spinning down neutron star was shown to be too under-luminous, by $\approx 4$ orders of magnitude, to explain the pulsed emission. Alternative accretion models were quickly ruled out, since there are none of the usual accretion hallmarks, like flickering in the light curves or complex line emission coming from mass transfer and accretion. Furthermore, the X-ray luminosity is 100th of that expected for an accreting white dwarf (i.e. a CV). 

Confirmation of the dipole emission model for AR Sco followed from polarimetric observations which revealed the system exhibits strongly pulsed linear polarization on both the spin and beat periods of the white dwarf, reaching 40\% \cite{Buckley2017, Potter2018}. The model proposed for AR Sco \cite{Marsh2016, Buckley2017, Potter2018} involves the interaction of the magnetic white dwarf’s magnetosphere with the red dwarf companion’s corona, trapping electrons which are accelerated to relativistic velocities, resulting in strongly beamed and polarized synchrotron emission. This general picture is supported by subsequent optical spectroscopy and Doppler tomography, which show that the emission lines originate close to the red dwarf star companion \cite{Garnavich2019, Pelisoli2022}, in a region periodically energized by the rotating white dwarf magnetosphere. Similarly, X-ray observations with XMM-Newton supported the same coronal injection model \cite{Takata2017}. Therefore, despite the pulsing behaviour that led to AR Sco being dubbed a white dwarf pulsar, the mechanism behind the emission is unlike the one for neutron star pulsars, and relies on binary interaction.

The apparent uniqueness of AR Sco amongst thousands of known white dwarf -red dwarf binaries was an aspect that puzzled Tom – was it a fluke or were there other systems like it, waiting to be discovered? Together with then newly hired postdoc Ingrid Pelisoli, a targeted search for white dwarf ``pulsars'' like AR Sco was started in early 2021. The search took advantage of the flagship {\it Gaia} mission, which indicated that AR Sco was in a relatively underpopulated region of the colour-magnitude diagram. This was combined with a search for infrared variability – another uncommon feature of AR Sco when compared to the rest of the white dwarf population – to identify just over 50 candidate white dwarf pulsars.

Follow-up of these systems was carried out with the fast photometers ULTRACAM, HiPERCAM and ULTRASPEC, in whose design and operation Tom played an important role (see the next section). Finally, in late April 2022, Matthew Green – one of Tom’s former PhD students – observed candidate J191213.72-441045.1 (henceforth J1912-4410) during a night of Guaranteed Time Observations (GTO) and immediately flagged it as a promising candidate as it showed clear pulses with a period around 5 minutes. On the next night, Ingrid called in a favour with collaborators observing with the Southern Astrophysical Research (SOAR) Telescope and obtained an ID spectrum. The spectrum turned out to show features remarkably similar to those seen in AR Sco: narrow Balmer emission lines over a blue continuum, with visible molecular absorption features from a red dwarf. This \textit{fantastic} discovery (Tom’s words) motivated further follow-up.

J1912-4410 was immediately the subject of successful proposals to X-shooter – for time-resolved spectroscopy covering a possible orbital period of 4 hours, identified in data from TESS – and MeerKAT (led by Patrick Woudt) – to search for radio pulses like those seen for AR Sco. Tom also reached out to long-term collaborator David Buckley for photopolarimetric observations with the High speed Photo-Polarimeter (HIPPO) at the South African Astronomical Observatory (SAAO). Coincidentally, the same target had recently been brought to David’s attention by Axel Schwope, as it had been detected as an X-ray source by the eRosita mission. Axel subsequently led follow up observations with XMM-Newton to further characterise the X-ray emission \cite{Schwope2023}.

The large collection of datasets above confirmed that indeed J1912-4410 was a second white dwarf pulsar: it clearly showed periodic pulses with a 5.3-min period all the way from radios to X-rays, in a similar fashion to AR Sco. The optical polarimetric properties were also akin to AR Sco, with significant (4-12\%) linear polarisation and no circular polarisation, suggesting a potential synchrotron origin for the pulsed emission. The discovery was published in Nature Astronomy in June 2023 \cite{Pelisoli2023}, a few months after Tom’s untimely passing. Even though Tom did not see the final form of the manuscript, it was agreed between the authors that he should be the second author of the article, given his central role in the project. 

None of the other initially identified candidates turned out to be white dwarf pulsars, but Tom did witness the one discovery resulting from the targeted search and had an answer to his question \textit{is AR Sco unique?}

No new confirmed white dwarf pulsars have been published since the discovery of J1912-4410, but searches continue. The field has gained additional attention with the discovery that some long-period radio transients – ILT J1101+5521 \cite{deRuiter2024} and GLEAM-X J0704‑3 \cite{Hurley-Walker2024, Rodriguez2025} – are in fact white dwarf plus red dwarf binaries, suggesting a possible connection to white dwarf pulsars, although the white dwarf in these systems does not appear to be fast spinning and the radio emission is modulated only on the orbital period. With new long-period radio transients being discovered at a fast pace \cite{lpt1, lpt2, lpt3, lpt4, deRuiter2024, Hurley-Walker2024, lpt5, lpt5b}, Tom’s work on white dwarf pulsars is in the spotlight again and could play a crucial role in revealing the nature of these puzzling radio sources, demonstrating the legacy value of his work.

\subsection*{Instrumentation}
{\large Vik Dhillon \& Stuart Littlefair}
\vspace{3mm} 

Tom leaves behind him a considerable legacy in astronomical instrumentation, primarily in the form of the high-speed imaging photometers ULTRACAM, ULTRASPEC and HiPERCAM.

For years, we had been frustrated in our attempts to study short-period variable stars by the slow readout times of the CCD detectors used at the world's major observatories. So, during a meeting in Stratford-upon-Avon in 1998, we came up with the idea of building our own instrument, exploiting the new frame-transfer CCDs that had just become available. Our previous experience with the dual beam spectrometer ISIS on the William Herschel Telescope (WHT) had also introduced us to dichroic beamsplitters, and so we decided to employ these as well to make a multi-colour, high-speed camera. The resulting triple-beam instrument, ULTRACAM, was funded by the Particle Physics and Astronomy Research Council (PPARC) and saw first light on the WHT in 2002 \cite{dhillon07}. ULTRACAM was then moved to Chile, where it became the first instrument to use the Visitor Focus of the VLT in 2005, and it is now permanently mounted on the New Technology Teslecope (NTT). To date, ULTRACAM has been used for 869 nights on 4--8\,m-class telescopes over 43 semesters, resulting in over 250 refereed publications, including Tom's discovery of white-dwarf pulsars \cite{Marsh2016}.

Our attention then turned to high-speed spectroscopy, which is particularly difficult due to the readout noise of CCD detectors. Recognising the potential of EMCCDs in this respect, we were awarded funding by the EU in 2004 to build ULTRASPEC, an EMCCD-based camera which saw first light on the EFOSC spectrograph of the ESO 3.6m in 2006, and the NTT in 2009. By removing readout noise we effectively turned the 3.6m telescope into a 6.3m telescope for high speed spectroscopy. This was the first time that an EMCCD had been used for astronomical spectroscopy on a large-aperture telescope, and its success led to an application to the ERC to build a larger-format version of ULTRASPEC for use on VLT/FORS, and an application to ESO to build an optical spectrograph for the ELT using EMCCDs (see \cite{tulloch11}). Unfortunately, both applications failed.

Although the case for low-noise detectors on astronomical spectrographs remains a strong one, we decided to return to imaging. First, we converted ULTRASPEC into an imager and mounted it permanently on the 2.4m Thai National Telescope (TNT). ULTRASPEC has been the main instrument on the TNT since first light in 2013 \cite{dhillon14}, resulting in over 70 refereed papers, including the discovery of minute-duration optical flares from an AT2018cow-like transient, implying the embedded energy source is a magnetar or accreting black hole \cite{ho23}. 

\begin{figure}[h!]
\centering
\includegraphics[width=0.85\textwidth]{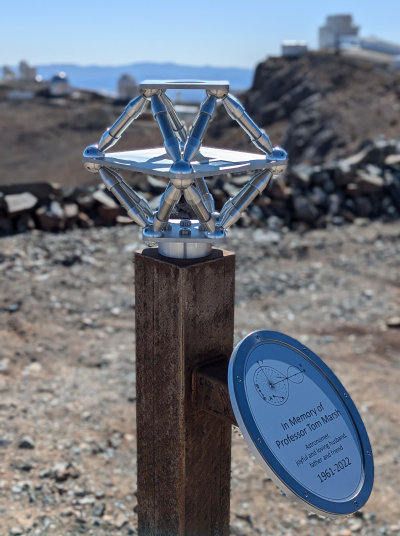}
\caption{A memorial for Tom has recently been installed on La Silla. On the top of the post is a scale model of ULTRACAM, and at the front a plaque with some words of remembrance and one of Tom's
well known Doppler Tomography diagrams \cite{MarshHorne1988}. We hope visitors to the site who knew Tom may take some comfort on their next visit to La Silla by walking up to the memorial to remember him.
\label{fig:memorial}}
\end{figure}

With ULTRASPEC commissioned on the TNT, our attention turned to HiPERCAM -- a new imager which built on the heritage of ULTRACAM and ULTRASPEC and which would give us multi-colour instruments in each hemisphere. We obtained funding from the ERC to build HiPERCAM in 2014, and the instrument saw first light on the GTC in 2018 \cite{dhillon21}. Given that it is mounted on the world's largest optical telescope, and that it images simultaneously in 5 colours covering the whole optical spectrum with essentially no dead time between exposures, HiPERCAM is arguably the world's most sensitive optical camera. Furthermore, in 2023 we commissioned a new rotator for HiPERCAM which means it is now permanently mounted at its own focus on the GTC, significantly increasing the amount of available observing time.  We are also currently designing a polarimetry module for HiPERCAM, which Tom was really looking forward to use. With approximately 45 refereed papers from 759 hours of telescope time, 7 of them in Nature/Science (including the discovery of the rings around Quaoar \cite{morgado23}), HiPERCAM has now become the most productive instrument on the GTC -- a fitting tribute to Tom (see also Fig.~\ref{fig:memorial}).

\clearpage

\subsection*{Teaching}
{\large Daniel Bayliss}
\vspace{3mm} 

In addition to being a brilliant researcher, Tom was also an enthusiastic and passionate lecturer.  Over his career he taught many undergraduate courses on both general physics and astrophysics.  Many of the lecture modules still taught today at the University of Warwick were in large part created by Tom.  In particular Tom put together the 1st year course ``Astronomy" (PX158), a core course for all Physics students covering an introduction to Astronomy.  Rather than relying on a textbook, Tom wrote a comprehensive introduction to astronomy complete with diagrams and slides.  The questions that Tom prepared for the Astronomy course reflected Tom's curiosity and creativity.  As just a flavour of these questions, here is one that Tom devised for 1st-year students relating to Stonehenge and the solstices:
\begin{quote}
\textit{
It is sometimes said that the latitude of Stonehenge (51.43$^{\circ}$) is significant because the four directions defined by the Sun when it rises and sets at the summer and winter solstices are at 90$^{\circ}$ to each other. Assess the truth of this claim.
}
\end{quote}

\begin{figure}[h!]
\centering
\includegraphics[width=0.8\textwidth]{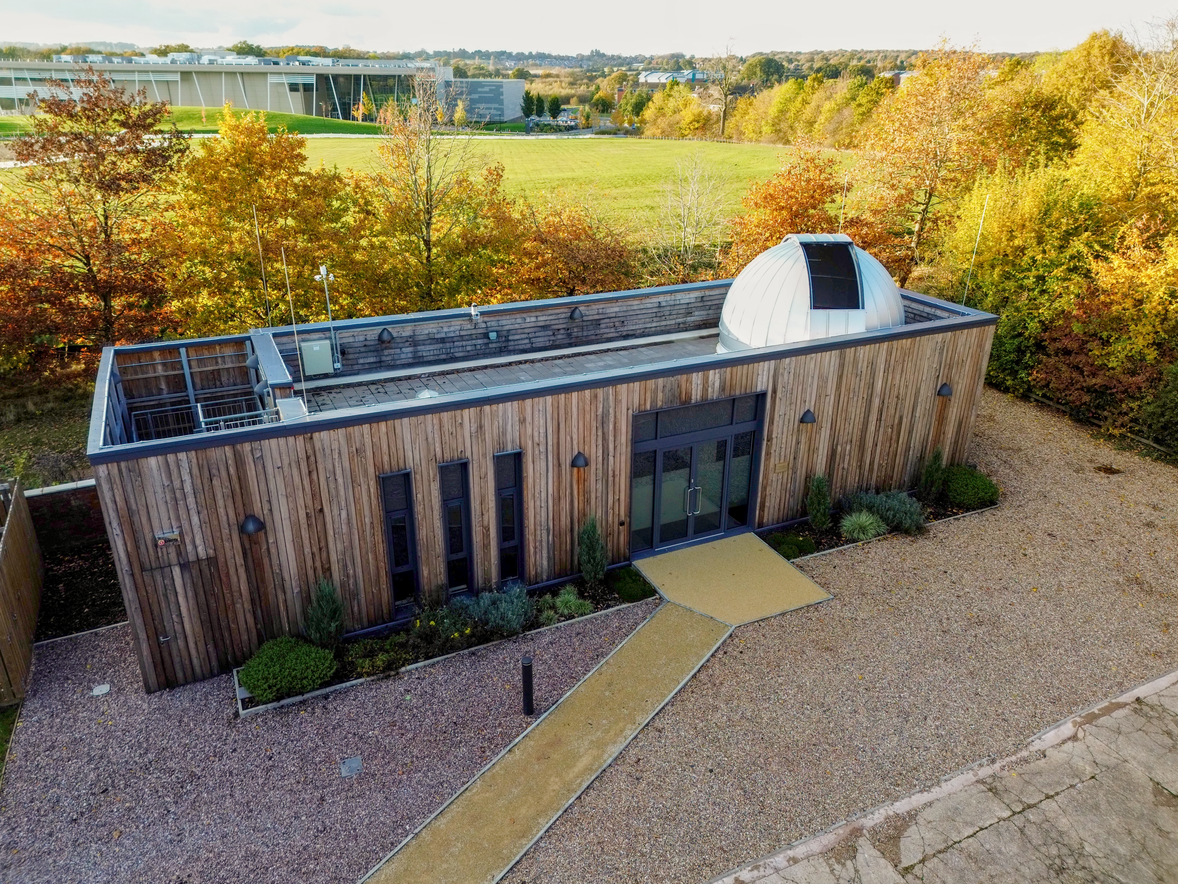}
\caption{The Marsh Observatory at the University of Warwick.
\label{fig:observatory}}
\end{figure}

The legacy of Tom's teaching also extends to the labs, where Tom's practical side shone through. Tom helped devise and implement labs that again continue on to this day at the University of Warwick. In tribute to Tom the newly built on-campus observatory, which is mainly used for undergraduate teaching and outreach, has been named The Marsh Observatory (see Fig.~\ref{fig:observatory}).

\newpage
\section{The History of Warwick Astro}
\label{sec2}
{\large Boris G\"ansicke}
\vspace{3mm}

The history of the Warwick Astronomy \& Astrophysics groups began sometime in January 2001, when the Physics Department at Warwick discussed an expansion beyond their strong focus on condensed matter research. A job ad at the professor level went out at the end of October 2002, announcing the opportunity to start a new research group with initially one additional lecturer. Tom, at that time a Reader at the University of Southampton, was encouraged to apply and interviewed by the end of March 2003. Warwick offered the job to Tom on April 1, 2003 (yes, Fools Day) \ldots and he turned it down, arguing that he would like to bring along Boris G\"ansicke from day one, but ultimately needed three lecturers to make a new Astronomy \& Astrophysics group viable. Because both Tom and Boris were on fellowships at Southampton, the University agreed to offer two lectureships with an option on a third one. Tom accepted the upgraded offer. His first day in the office was August 29, 2003, and Boris joined him November~1. 

The group was given three adjacent offices in the ground floor of the Physical Sciences building, and Tom chose office PS007 (for semi-obvious reasons). Besides the two academic staff, the group consisted of Tom’s PhD student Susana Barros (now research staff at the Instituto de Astrof\'isica, Portugal) and Boris’ student Amornrat Aungwerojwit (now faculty at Naresuan University, Thailand) and postdoc Pablo Rodr\'guez-Gil (now faculty at the University of La Laguna, Spain). Their research focused on compact binaries and accretion physics, though Tom kept the option of expanding beyond those areas in the future. J\"orn Wilms, an expert in X-ray astronomy, X-ray binaries and AGN joined the group as a lecturer in 2004, followed in 2005 by Peter Wheatley, who had a strong background in compact binaries, but was transitioning into exoplanet science. J\"orn left the group in 2006, and was replaced by Danny Steeghs, working also on compact binaries and accretion physics, and Andrew Levan, studying gamma ray bursts. The two were so closely ranked in the job interview that Tom thought it was impossible to distinguish between them, and after a call to the Vice Chancellor, the decision was made to hire both. The group was very close-knit, with one of its members remembering fondly ``\textit{I enjoyed the main mode of communications in those early years where we were all in one row of offices and Tom would just shout out stuff which we could all hear}''. 

The group remained stable for a number of years, until the opportunity to hire another lecturer came along in 2010. The decision whether to remain focused on the existing topics, or to branch out further was rather heatedly discussed, and ended with the appointment of Elizabeth Stanway in 2011, working on high-redshift galaxies. Around the same time, Warwick was planning a strategic expansion, and started an initiative to hire “star professors”. By name, those nicely fitted into the remit of the Astronomy \& Astrophysics group. Don Pollacco, then professor at Queen's University Belfast, was approached and eventually hired in 2012. Don’s appointment massively increased the focus of instrumentation at Warwick, which so far was limited to Tom’s involvement in ULTRACAM and its siblings. Don had built the very successful SuperWASP project, searching for transiting planets, and the concept of small aperture telescope arrays led to the development of the ground-based NGTS and GOTO and the PLATO mission of the European Space Agency. 

Jumping on an opportunity to hire a theoretical physicist, the group approached Pier-Emmanuel Tremblay, then at STScI and an expert in white dwarf atmospheres, who was hired as a lecturer in 2015. 

In 2017, the group argued that the exoplanet side required strengthening to remain competitive, and underwent a phase of inflation, with the hires of Farzana Meru (theory of protoplanetary discs), Dan Bayliss (exoplanet transits), Matteo Brogi (exoplanet atmospheres), Dimitri Veras (theory of the evolution of planetary systems), David Armstrong (exoplanets) and Grant Kennedy (debris discs). This growth surprised even the most optimistic members of the group, and was fuelled by the fact that most of the new staff members were holding long-term fellowships. Heather Cegla (stellar surfaces and activity) was appointed in 2020.

Finally, post-Covid, the group went through some consolidation, appointing Deanne Coppejans (transients and compact binaries), Joe Lyman (extragalactic transients), Ingrid Pelisoli (white dwarf and hot subdwarf binaries), Siddarth Gandhi (exoplanet atmospheres, replacing Matteo Brogi) and Rebecca Nealon (theory of accretion discs). 

Under Tom’s visionary leadership, Astronomy \& Astrophysics at Warwick grew within 20 years from just five members to one of the largest groups in the UK. At the time of writing, the group consists of 17 faculty members, eight research and teaching staff, 28 postdocs and 43 PhD and MSc students. 

\begin{figure}[h!]
\centering
\includegraphics[width=1\textwidth]{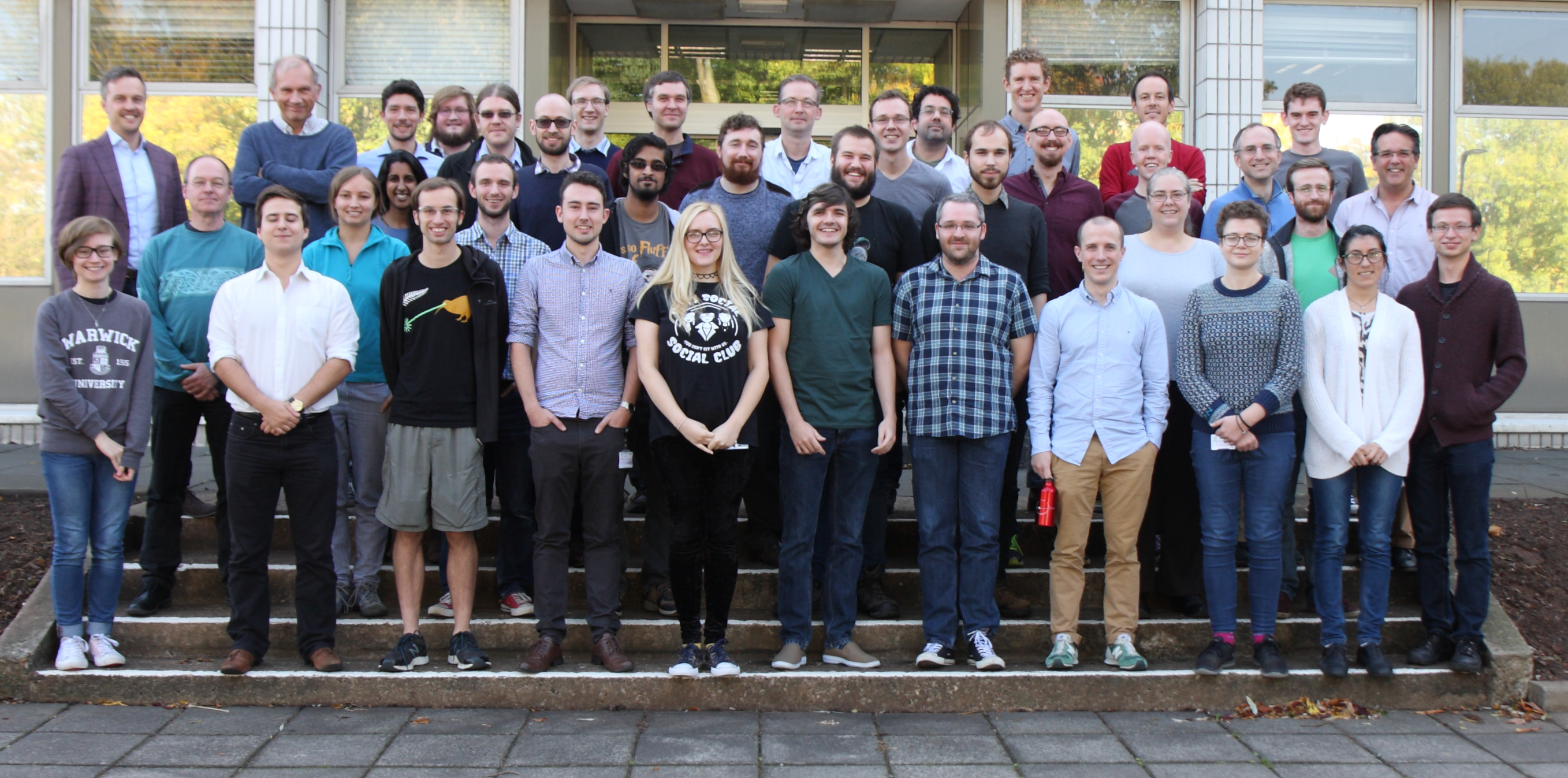}
\includegraphics[width=1\textwidth]{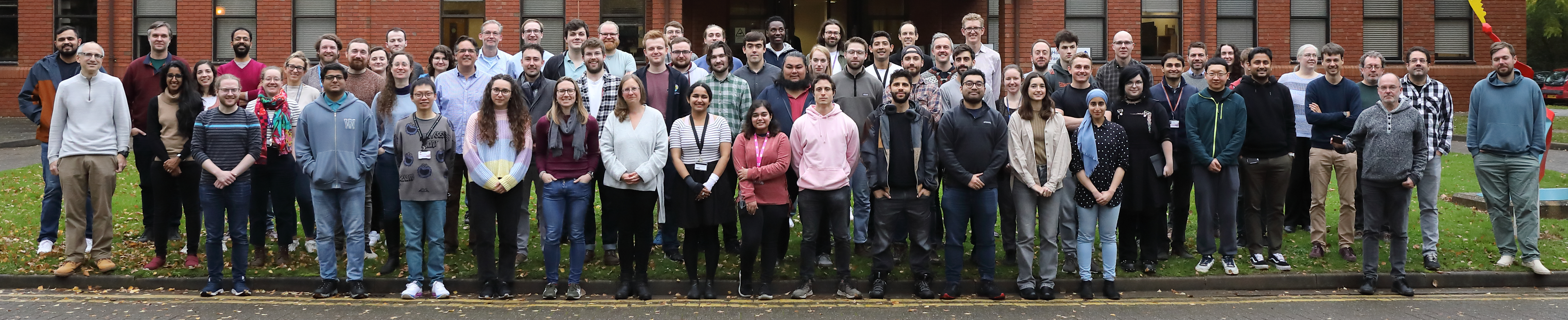}
\caption{The Astronomy \& Astrophysics group in 2018 (top, Tom is second-left in the top-most row) and 2024.
\label{fig:group}}
\end{figure}

\newpage

\bibliographystyle{unsrtnat}
\bibliography{references}

\end{document}